\documentclass[a4paper, 11pt]{article}

\pdfoutput=1

\usepackage{graphicx}
\usepackage{ subfig}
\usepackage{color}

\def\b{\begin{eqnarray}}
\def\e{\end{eqnarray}}
\def\g{\gamma}
\def\n{\noindent}
\begin{document}

\begin{center}

{\huge \textbf{Cyclic Universe with an Inflationary \vskip0.3cm Phase from a Cosmological  Model with \vskip0.5cm Real Gas Quintessence}}

\vspace {10mm}
\noindent
{\large \bf Rossen I. Ivanov and Emil M. Prodanov} \vskip.5cm
{\it School of Mathematical Sciences, Dublin Institute of Technology,
Ireland,} \vskip.1cm
{\it E-Mails: rossen.ivanov@dit.ie, emil.prodanov@dit.ie} \\
\vskip1cm
\end{center}

\vskip4cm
\begin{abstract}
\n
Phase-plane stability analysis of a dynamical system describing the Universe as a two-fraction fluid containing baryonic dust and real virial gas
quintessence is presented. Existence of a stable periodic solution experiencing inflationary periods is shown. A van der Waals quintessence
model is revisited and cyclic Universe solution again found.
\end{abstract}

\newpage

\section{Introduction}
There has been an extensive search for new forms of matter, energy and possible new physics beyond the four fundamental forces described by
the Standard Model and General Relativity to conform with the present day observational picture in cosmology and, in particular, to account for
the cosmic acceleration \cite{1}. A fifth element, {\it quinta essentia}, in addition to baryons, leptons, radiation, and cold dark matter, has been
sought, typically, in the form of cosmological constant, scalar fields, hybrid models involving both fluid and scalar field, modification of gravity,
quantum gravity models, etc --- see, for example, \cite{2}, \cite{3}, \cite{4}, \cite{5}, \cite{6}, \cite{7}, \cite{8}, \cite{9}, \cite{10}  and the references,
therein. \\
A classical physics model with a two-phase fluid, described by a van der Waals equation of state, and with no need of introduction of scalar fields,
has also been studied \cite{cap1}, \cite{cap2}, \cite{cap3} as quintessence. In this model, negative absolute temperature is needed to explain
accelerated expansion if the present day energy density of the dark matter (described as the van der Waals fluid) is much smaller than the van der
Waals critical density. \\
This paper presents another classical physics model describing the Universe as an infinite, flat, two-fraction mixture of baryonic dust and a real gas
with equation of state derived from the virial expansion. Dynamical phase-plane analysis (with Hubble's parameter $H$ and density of the real gas
$\rho$ as dynamical variables) of such model shows that there is initial data leading to a cyclic Universe solution that goes through an inflationary
phase in each cycle, together with open trajectories in the phase plane that may or may not pass through regions characterized by inflation. The
temperature in this model is also treated as a parameter. However, this parameter that does not have to take negative values. It is shown that, as the
Universe cools down, the inflationary region on the phase plane decreases and eventually disappears in the limit $T \to 0$. The cosmological
model presented here also does not exhibit an endless sequence of cycles of expansion and contraction. The trajectory of the Universe on the
phase place is, in first approximation, an ellipse and the frequency of oscillations decreases, while the ratio of of its axes decreases as the
Universe is cooling. Eventually periodicity is lost as well. \\
It is essential to point out that the particle interaction of this model has been described as two-particle interaction between indestructible ``hard
spheres" surrounded by finite potential wells --- a regularized form of a classical real gas interaction which is slightly attractive at long distances
and strongly repulsive at short ones. The van der Waals model is also revisited in the light of seeking similar cyclic solution (which is found for
negative temperatures). A stable-node equilibrium solution for positive temperatures is also present.

\section{Cosmological Setup}
The setting for the analysis of the two-fraction Universe is the Friedmann-Robertson-Walker-Lema\^itre (FRWL) cosmology \cite{frwl} with metric:
\b
ds^2 = g_{\mu \nu} dx^\mu dx^\nu =
c^2 dt^2 - a^2(t) \Bigl[\frac{dr^2}{1 - kr^2} + r^2 (d \theta^2 + \sin^2 \theta \, d \phi^2) \Bigr],
\e
where $a(t)$ is the scale factor of the Universe and $k$ is the spatial curvature parameter. \\
Einstein's equations are:
\b
G_{\mu \nu} + \Lambda g_{\mu \nu} = \kappa T_{\mu \nu},
\e
where $\kappa = 8 \pi G/c^4$ (in Planck units, $\kappa = 1$) and the matter energy-momentum tensor $T_{\mu \nu}$ is given by:
\b
T_{\mu \nu} = (\tilde{\rho} \, c^2 + \tilde{p})c^2 \, \delta_\mu^0 \, \delta_\nu^0 - \tilde{p} \, g_{\mu \nu} \, ,
\e
with $\tilde{\rho}$ and $ \tilde{p}$ --- the cumulative density and pressure for both fractions. \\
Friedmann equations are \cite{friedmann}:
\b
\label{fr1}
\ddot{a} & = & - \frac{4 \pi G}{3} (\tilde{\rho} + \frac{3 \tilde{p} }{c^2})a + \frac{1}{3} \Lambda c^2 a  \\
\label{fr2}
\dot{a}^2 & = &  \frac{8 \pi G}{3} \tilde{\rho} a^2  + \frac{1}{3} \Lambda c^2 a^2 - c^2 k.
\e
Only the case of flat spatial three-sections ($k=0$) and without cosmological constant (i.e. $\Lambda = 0$) will be of interest. \\
The Hubble parameter $H$ is $H = \dot{a}/a$ and this will be one of the dynamical variables. In terms of $H$, in Planck units, with $k=0$ and
$\Lambda =0$, Friedmann equations (\ref{fr1})--(\ref{fr2}) become:
\b
\label{h1}
H^2 & = & \frac{1}{3} (\rho_b + \rho), \\
\label{h2}
\dot{H}& = & - \frac{1}{2} (\rho_b + \rho + p).
\e
The continuity equation for the real gas
\b
\label{cont}
\dot{\rho} + (\rho + p) \frac{3 \dot{a}}{a} = 0
\e
becomes
\b
\label{h3}
\dot{\rho} + 3H(\rho + p) = 0,
\e
while the continuity equation for the pressureless baryonic dust is:
\b
\label{h4}
\dot{\rho_b} + 3H \rho_b  = 0.
\e
Differentiating (\ref{h1}) with respect to time and substituting into it $\dot{H}$ from (\ref{h2}),
$\dot{\rho}$ from (\ref{h3}) and $\dot{\rho_b}$ from (\ref{h4}), leads to an identity. Thus, equation (\ref{h1}) is just an integral of equations (\ref{h2}),
(\ref{h3}), and (\ref{h4}).  As it can be obtained from equations (\ref{h1}), (\ref{h2}) and (\ref{h3}), equation (\ref{h4}) will be droppped. \\
Expressing the baryonic energy density $\rho_b$ from equation (\ref{h1}) and substituting it into equation (\ref{h2}) gives the dynamical equation
\b
\label{hash}
\dot{H} = - \frac{3}{2} H^2 - \frac{1}{2} p.
\e
The other dynamical equation is (\ref{h3}):
\b
\label{rho}
\dot{\rho} = - 3H(\rho + p),
\e
with $\rho$ being the second dynamical variable.

\section{Real Virial Gas Quintessence}
The virial expansion relates the pressure $p$ of a real gas to the particle number $N$, the temperature $T$ and the volume $V$ of the gas
 \cite{mandl}:
 \b
 \label{vir}
 p = \frac{Nk_BT}{V} \Bigl[ 1 + \frac{N}{V} F(T) + \Bigl( \frac{N}{V} \Bigr)^2 G(T) + \cdots \Bigr].
 \e
The ``departure" from ideal gas ($p = Nk_BT/V$) is due to the interactions between the gas particles which, also, no longer have zero volumes.
These are included in the above expression via the correction term $F(T)$, describing the two-particle interactions; the correction term $G(T)$, for
the three-particle interactions, and so forth. \\
The correction term $F(T)$ is \cite{mandl}:
 \b
 F(T) = 2 \pi \int\limits_{0}^{\infty} \Bigl[1 - e^{-\frac{V(r)}{k_BT}} \Bigr] \, r^2 dr,
 \e
 where $V(r)$ is the two-particle interaction potential. \\
For the purposes of this work, considered here is a classical real gas model of particle interaction that never involves more than two particles
(ignoring collisions of 3 particles or more amounts to disregarding the term $G(T)$ and the subsequent terms in the virial expansion) and which is
slightly attractive at long distances and strongly repulsive at short ones (the dashed curve on Figure 1). Such interaction is often presented in a
regularized form as an ensemble of  identical impenetrable ``hard spheres" of radius $a$ (namely, two real gas particles not being able to get
closer to one another at distance between their centres smaller than $2a$), surrounded by square potential wells of width $ad$ ($d > 1$, $d=1$
corresponds to absence of well) and depth $- \epsilon$ (where $\epsilon > 0$). The regularized potential (the solid curve on Figure 1) of
the two-particle interaction is:
\b
 \label{potential}
 V(r) = \left\{
 \begin{array}{ll}
 \infty \, , & \mbox{$0 < r < a,$} \cr
 - \epsilon , &  \mbox{$a \le r \le ad, $}
 \cr  0 \, , & \mbox{$r > ad.$}
 \end{array}
 \right.
 \e
It is important to point out, that representing the interactions as a collection of ``hard spheres" only $(d = 1)$ is not sufficient for the analysis that
follows. \pagebreak Namely, the attractive part (the square well) is essential and thus $d > 1$.
\begin{center}
\vskip-1.4cm
\includegraphics[width=9cm]{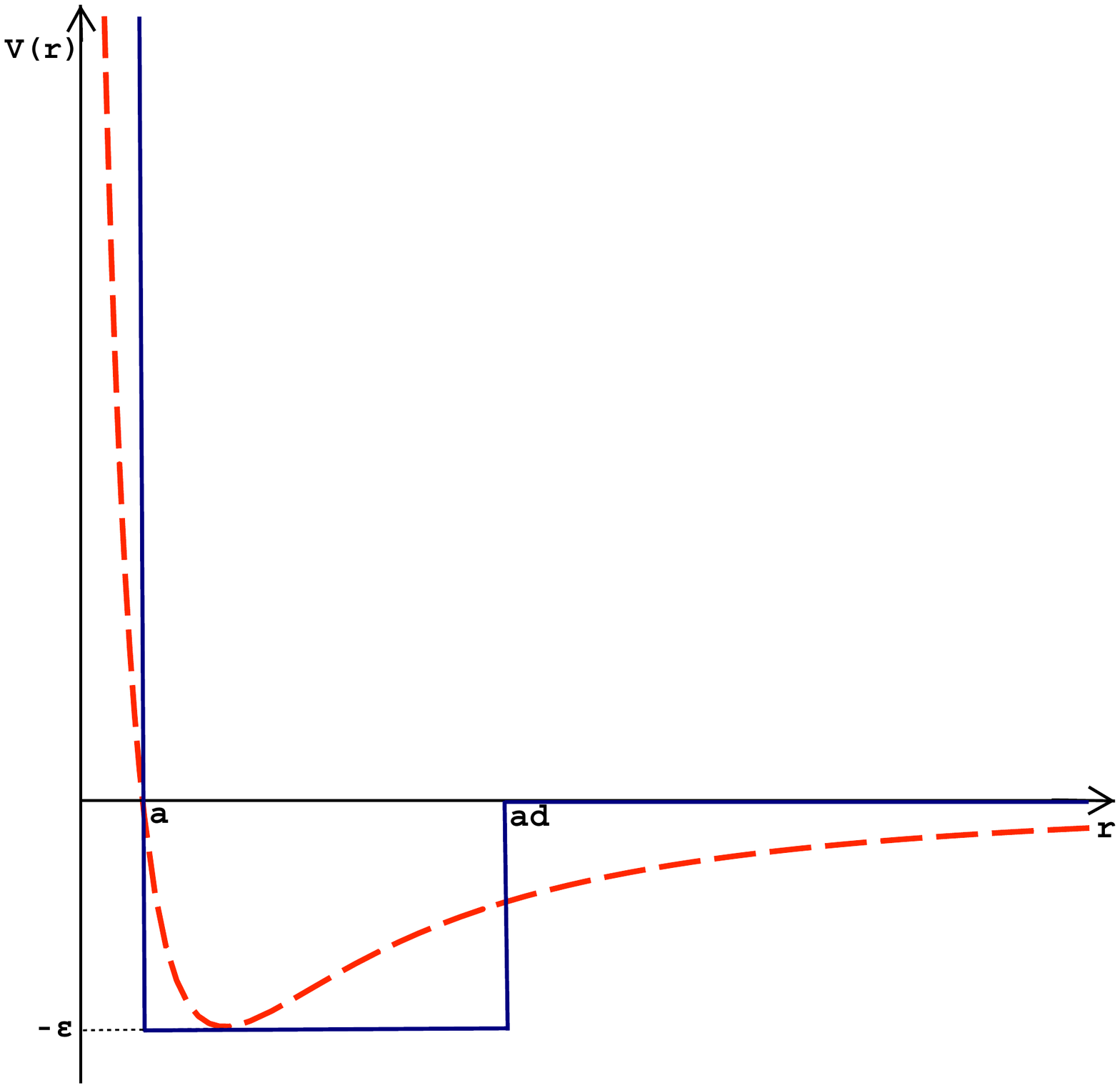}
\vskip-1.4cm
\noindent
\parbox{145mm}{\hskip0.84cm Figure 1:   \footnotesize The interaction potential (dashed curve) and its regularized form (solid curve).}
\vskip.4cm
\end{center}
\n
With the expansion of the exponent up to and including the term proportional to $1/(k_BT),$ the correction $F(T)$ becomes:
 \b
 F(T) = B - \frac{A}{k_BT}
 \e
 and the virial expansion (\ref{vir}) becomes the  ``van der Waals'' equation \cite{mandl}:
 \b
 \label{pvdw}
 p + \Bigl( \frac{N}{V} \Bigr)^2 A  =  \frac{Nk_BT}{V} \Bigl( 1 + \frac{N}{V} B \Bigr).
 \e
In the limit $N B / V \to 0$, this equation reduces to the usual van der Waals equation (which was presented before the real gas virial expansion
form (\ref{vir}) and is not ``derivable" from it) \cite{mandl}:
 \b
 \label{vdw}
 \Bigl[ p + \Bigl( \frac{N}{V} \Bigr)^2 A \Bigr]  \Bigl( 1 - \frac{N}{V} B \Bigr) =  \frac{Nk_BT}{V}  \, .
 \e
Phase-plane stability analysis of a dynamical system describing the Universe as a two-fraction fluid containing baryonic dust  with energy density
$\rho_b$ and pressure $p_b = 0$, and van der Waals gas quintessence  will be discussed later in this paper. Firstly, this analysis will be done for
a quintessence of real gas whose two-particle interaction potential is described by (\ref{potential}). In Planck units ($k_B=1$), the correction term $F(T)$ for such gas is given by:
\b
F(T) = 2 \pi \biggl[ \int\limits_{0}^{a} r^2 dr  \, + \, \int\limits_{a}^{ad} r^2 \Bigl( 1 - e^{\frac{\epsilon}{T}} \Bigr) dr \biggr] = \frac{2 \pi a^3}{3}
[ 1 +   (1 - e^{\frac{\epsilon}{T}}) (d^3 - 1)]  = - \alpha z(T),
\e
where  $z(T) = (e^{\frac{\epsilon}{T}} - 1) (d^3 - 1) - 1$ and $\alpha = (2/3) \pi a^3$. \\
For gases containing spherical molecules only, with a square well interaction potential (\ref{potential}) between them, the next virial
coefficient $G(T)$ is also known \cite{kih} --- it is a term cubic in $e^{\frac{\epsilon}{T}} - 1$. However, the inclusion of such asymptotically smaller term
does not significantly alter the physics of the model presented here and for that purpose three-particle interactions or interactions involving collisions
of four or more particles will be disregarded.  \\
In terms of the gas density, the real gas equation of state is:
\b
\label{eos}
p = T \rho [1 - \alpha \rho z(T)].
\e
In the analysis, $\alpha, \epsilon,$ and $d$ will be the parameters of the model. The temperature $T$ will also be treated as a parameter, but one
that will be allowed to vary, and the different values of this parameter, would single out different stages of the ``evolution" of the trajectories in the
phase-plane of the dynamical system.
Upon substitution of the equation of state (\ref{eos}), the dynamical system becomes:
\b
\label{dyn1a}
\dot{\rho} & = & - 3 H \rho \,  [1 + T - \alpha \rho T z(T)  ] \equiv f_1(\rho, H), \\
\label{dyn1b}
\dot{H} & = & - \frac{3}{2} H^2 - \frac{1}{2} T \rho [ 1- \alpha \rho z(T) ]  \equiv f_2(\rho, H).
\e
The first observation is that the origin of the phase-plane is an equilibrium point --- at this point, $\dot{\rho} = 0 = \dot{H}$. Equilibrium points are
standardly denoted with stars and the equilibrium point at the origin is therefore:  $(\rho^\ast_1 = 0, \,\, H^\ast_1 = 0).$ \\
The next observation is that for equilibrium points to occur away from the origin of the phase-plane, the pressure $p$ must not be positive. This is
visible from equation (\ref{hash}) --- at an equilibrium point, $\dot{H} = 0$, thus, if $H^\ast \ne 0$ at that point, then (\ref{hash}) is consistent only
when $p < 0$ ($H$ has to be real). As $\alpha$ is positive (it is one half of the volume of the ``hard spheres") and the density
$\rho$ is also positive, the pressure $p$ is negative when:
\b
\label{neg}
1 - \alpha \rho z(T)   < 0.
\e
Therefore, equilibrium points, for which $H^\ast \ne 0$, can occur only for values of the density $\rho$  greater than
\b
\label{rmin}
\rho_{\mbox{\tiny min}} = \frac{1}{\alpha z(T)} = \frac{1}{ \alpha [(e^{\frac{\epsilon}{T}} - 1) (d^3 - 1) - 1 ]  }.
\e
Clearly, $z(T) = (e^{\frac{\epsilon}{T}} - 1) (d^3 - 1) - 1$ must be positive in order $\rho_{\mbox{\tiny min}}$ to be positive. This puts an upper limit
on the temperature below which equilibrium points, different from the origin of the phase-plane, exist:
\b
T_{\mbox{\tiny max}} = \frac{\epsilon}{\ln d^3 - \ln (d^3 -1)}.
\e
This temperature is known from the theory of molecular gases as Boyle temperature $T_B$ \cite{kih}. It is called Boyle temperature because
at  around this temperature, Boyle's law holds best \cite{kih}. This is the temperature at which the first virial coefficient $F(T)$ --- or, for our purposes,
$z(T)$  --- changes sign. In our analysis, according to the inequality (\ref{neg}), the temperature of the virial gas must stay at all times bellow the Boyle
temperature. This is the reality condition for the Hubble parameter $H$, if equilibrium points are to exist away from the origin of the phase plane. \\
For example, the Boyle temperature is 123 K for Ne, 410 K for A,  594 K for Kr, 772 K for Xe \cite{kih}. \\
Relative to the Boyle temperature $T_B$, one can define ``van der Waals molecular volume" $v_B$ as follows:
\b
v_B = \Bigl[ T \frac{dF(T)}{dT}\Bigr]_{T=T_B}
\e
--- as it is equal to the term interpreted as molecular volume in the van der Waals equation of state \cite{kih}. The relationship between $\alpha$ and
$v_B$ is \cite{kih}:
\b
\frac{v_B}{\alpha} = d^3 [ \ln d^3 - \ln (d^3-1)].
\e
The van der Waals volume $v_B$ is 35.5 \AA$^3$ for Ne, 67.4 \AA$^3$ for A, 78.5 \AA$^3$ for Kr, 114 \AA$^3$ for Xe \cite {kih}. \\
For the rare gases, the value of $d$ is about 2.0 \cite{kih}. \\
It will be shown further in the analysis that a cyclic Universe scenario with an inflationary phase emerges within this set-up. Sufficient expanding phase
can be provided by choosing parameters of the model in such way that the Boyle temperature is sufficiently high. For example, Boyle temperature of
 the order of $10^{10}$ K can be achieved by increasing the range of the interaction by a factor of 1000, while keeping
$\epsilon$ fixed. Also, increasing $\epsilon$ increases linearly the Boyle temperature. \\
For a fixed Hubble parameter $H$ [i.e. $\dot{H} = 0$ in (\ref{dyn1b})],  the temperature $T$ increases monotonically with the density $\rho$  and tends to
the Boyle temperature as $\rho$ grows to infinity. The domain of the function $T(\rho)$ is from $\rho_{\mbox{\tiny min}}$ [see (\ref{rmin})] to $\infty$.
There is a minimum of the function outside this domain ($H$ is not a real number there): at point $\rho_{\mbox{\tiny min}}/2$ and the value of the
temperature at this point is the solution of the transcendental equation $3H^2 = T/[4 \alpha z(T)]$. \\
Returning to the phase-plane analysis, it can be easily seen that there is another equilibrium point for which
$\rho^\ast_2 = \rho_{\mbox{\tiny min }}, \,\, H^\ast_1 = 0$.
There are two more equilibrium points. From $\dot{\rho} = 0$, it is straightforward to find
\b
\label{r2}
\rho^\ast_3 \,\, = \,\, \frac{1+T}{\alpha T z(T)}
\,\, = \,\, \rho_{\mbox{\tiny min}} (1 + \frac{1}{T}).
\e
Substituting this into (\ref{dyn1b}) and requesting $\dot{H} = 0$, leads to:
\b
\label{h23}
H^\ast_{2,3} \,\, = \,\, \pm \,\, \sqrt {\frac{\rho^\ast_3}{3}} \,\,
=  \,\, \pm \,\, \sqrt {\frac{1}{3} \rho_{\mbox{\tiny min}} (1 + \frac{1}{T})}.
\e
The last two equilibrium points are therefore $(\rho^\ast_3 \, , H^\ast_{2})$ and $(\rho^\ast_3 \, , H^\ast_{3})$. \\
Next, the dynamical system is linearized near an equilibrium point  $(\rho^\ast, H^\ast)$:
\b
\label{lin_dyn2}
\dot{\rho} & \equiv & f_1(\rho, H)  \,\, =  \,\, \Bigl(\frac{\partial f_1}{\partial \rho}\Bigr)^* (\rho - \rho^\ast) + \Bigl(\frac{\partial f_1}{\partial H}\Bigr)^*
(H - H^\ast) + \ldots, \\
\label{lin_dyn1}
\dot{H}  & \equiv & f_2(\rho, H) \,\, =  \,\, \Bigl(\frac{\partial f_2}{\partial \rho}\Bigr)^*(\rho - \rho^\ast) + \Bigl(\frac{\partial f_2}{\partial H}\Bigr)^* (H - H^\ast) +
\ldots,
\e
where the stars on the derivatives indicate that they are taken at the equilibrium point $(\rho^\ast, H^\ast)$. In matrix form this can be written as:
\b
\label{ds1}
\frac{d}{dt} X(t) = L(\rho^\ast, H^\ast) \cdot X(t),
\e
where:
\b
\label{ds2}
X(t)= \left(
\begin{array}{c}
\rho(t) - \rho^\ast \cr \cr
H(t) - H^\ast
\end{array}
\right)
\mbox{ and } \,\,
L(\rho, H) = \left(
\begin{array}{cc}
\frac{\partial f_1}{\partial \rho} & \frac{\partial f_1}{\partial H} \cr \cr
\frac{\partial f_2}{\partial \rho} & \frac{\partial f_2}{\partial H}
\end{array}
\right).
\e
The stability matrix $L(\rho, H)$ is:
\b
\label{L}
L(\rho, H) = \left(
\begin{array}{cc}
- 3 H(1+T) + 6H \alpha T \rho z(T) & - 3 \rho [1 + T - \alpha T \rho z(T) ]
\cr
- \frac{T}{2} + \alpha T \rho z(T) & -3H
\end{array}
\right).
\e
To determine the type of an equilibrium point, the eigenvalues of the stability matrix should be determined at that point. \\
Firstly, at the origin $(\rho^\ast_1 = 0 \, , H^\ast_{1} = 0)$, the characteristic equation of the stability matrix $L(0, 0)$ is $\lambda^2 = 0$. Thus,
$\lambda = 0$ is a double eigenvalue and to analyze this equilibrium point and see if the origin is stable under perturbations, a small infinitesimal
perutrbation parameter $\delta > 0$ will be introduced in the second equation of the dynamical system (\ref{dyn1a})--(\ref{dyn1b}):
\b
\label{da}
\dot{\rho} & = & - 3 H \rho \,  [1 + T - T \alpha \rho z(T) ] , \\
\label{db}
\dot{H}  & = & - \frac{3}{2} H^2 + \frac{3}{2} \delta^2 - \frac{1}{2} T \rho [ 1- \alpha \rho z(T) ].
\e
The introduction of the $\delta$-term leads to a bifurcation of the original equilibrium point $(\rho^\ast_1 = 0 \, , H^\ast_{1} = 0)$ into a ``dipole" ot two
equilibrium points $P^+$ and $P^-$ \pagebreak with coordinates on the phase plane $(\rho^\ast_1 = 0 \, , H'^\ast_{1} = + \delta)$ and
$(\rho^\ast_1 = 0 \, , H''^\ast_{1} = -\delta),$ respectively (Figure 2).
\begin{center}
\vskip-1.4cm
\includegraphics[width=8.5cm]{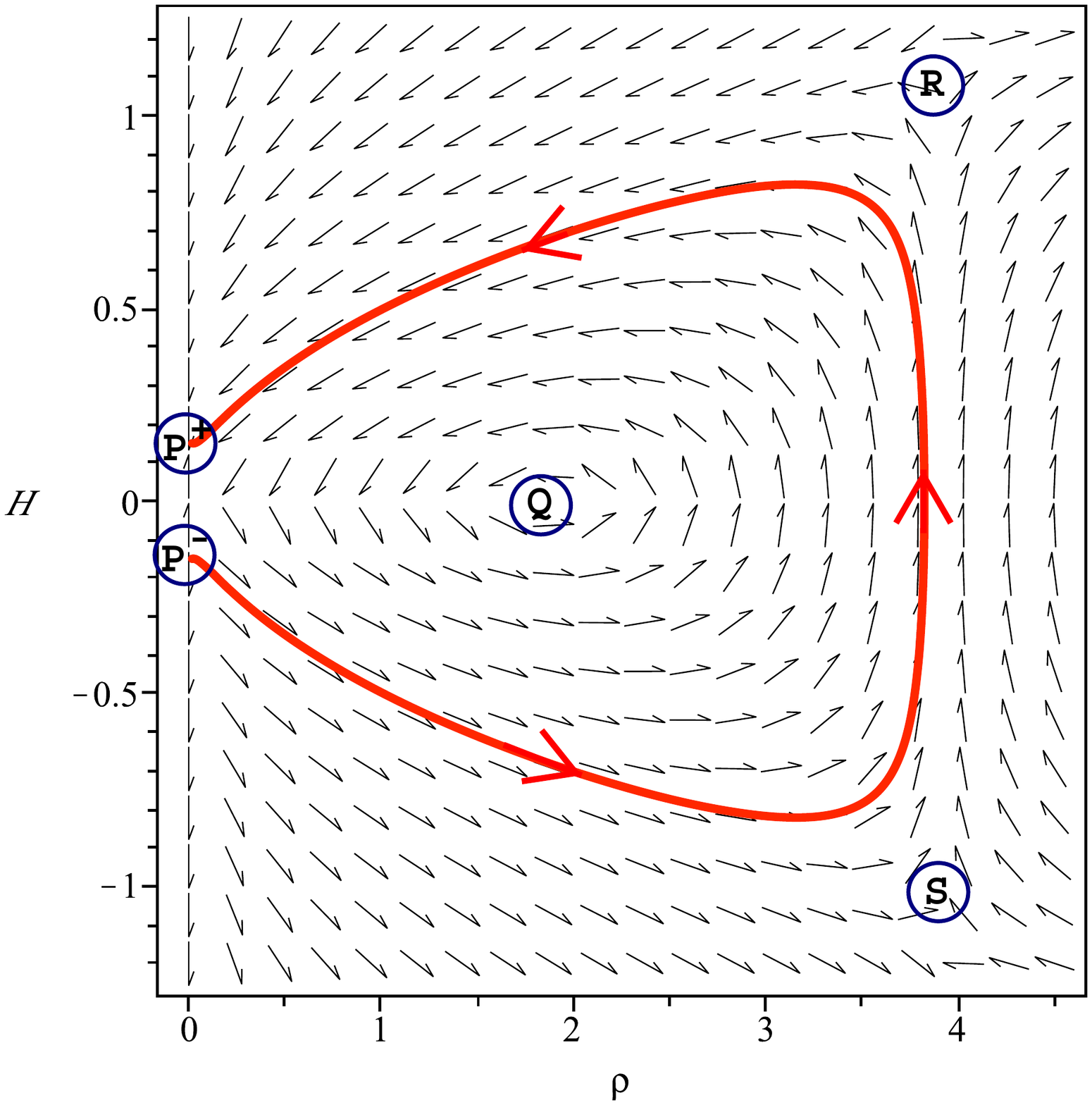}
\vskip-1.4cm
\noindent
\parbox{145mm}{Figure 2:   \footnotesize The phase-plane diagram of the perturbed system. Points $R$ and $S$ are saddles, $Q$ --- centre,point
$P^+$ --- stable node, point $P^-$ --- unstable node.}
\vskip.4cm
\end{center}
\n
At the upper point of the ``dipole", both eigenvalues are negative: $\lambda_1 = - 3 \delta (1 + T) < 0$ and  $\lambda_2 = - 3 \delta < 0.$ This point
therefore corresponds to a {\it stable node}. At the lower point of the ``dipole", both eigenvalues change only their signs. This is an {\it unstable
node} as both eigenvalues are positive. Upon taking the limit $\delta \to 0$, the stable and the unstable node coalesce and thus the origin
$(\rho^\ast_1 = 0 \, , H^\ast_{1} = 0)$ will attract trajectories in the phase-plane, as shown on Figure 2. \\
Next, at the equilibrium point $R$ with coordinates $(\rho^\ast_3 \, , H^\ast_{2})$ (determined in (\ref{r2})--(\ref{h23}) above), the eigenvalues of
$L(\rho^\ast_3 \, , H^\ast_{2})$ are:
\b
\lambda_1 & = & \sqrt{3 \rho^\ast_3} \,\,\, (1 + T) > 0, \\
\lambda_2 & = & - \sqrt{3 \rho^\ast_3} < 0.
\e
Therefore, as the eigenvalues have opposite signs, this equilibrium point is a {\it saddle point}. \\
At the equilibrium point $S$ with coordinates $(\rho^\ast_3 \, , H^\ast_{3})$ (see (\ref{r2})--(\ref{h23}) again), the eigenvalues only change their
signs. Thus, point $(\rho^\ast_3 \, , H^\ast_{3})$ is also a {\it saddle point} --- the eigenvalues again have opposite signs. \\
Finally, at the equilibrium point  $Q$ with coordinates $(\rho^\ast_2 = \rho_{\mbox{\tiny min}} \, , H^\ast_{1} = 0)$, the eigenvalues are purely
imaginary --- the characteristic equation is:
\b
\label{i}
\lambda^2 = - \frac{3T}{2 \alpha z(T)  }.
\e
This corresponds to a {\it centre}. The trajectories are closed curves, corresponding to a cyclic Universe. To analyze what happens and how the
temperature parameter affects these curves, it would be helpful to expand the dynamical variables $\rho$ and $H$ near the equilibrium point
$(\rho^\ast_2 = \rho_{\mbox{\tiny min}} \, , H^\ast_{1} = 0)$, namely: $\rho = \rho_{\mbox{\tiny min}} + r$ and $H = 0 +h$ .
The dynamical system (\ref{dyn1a})--(\ref{dyn1b}) becomes:
\b
\dot{r} & = &  - \frac{3}{\alpha z(T) } h + 3(T-1) h r + 3 \alpha T z(T) h r^2,  \\
\nonumber \\
\dot{h} & = & - \frac{3}{2} h^2 + \frac{1}{2}T r + \frac{1}{2} \alpha T z(T) r^2.
\e
\begin{center}
\vskip-1.4cm
\includegraphics[width=8.5cm]{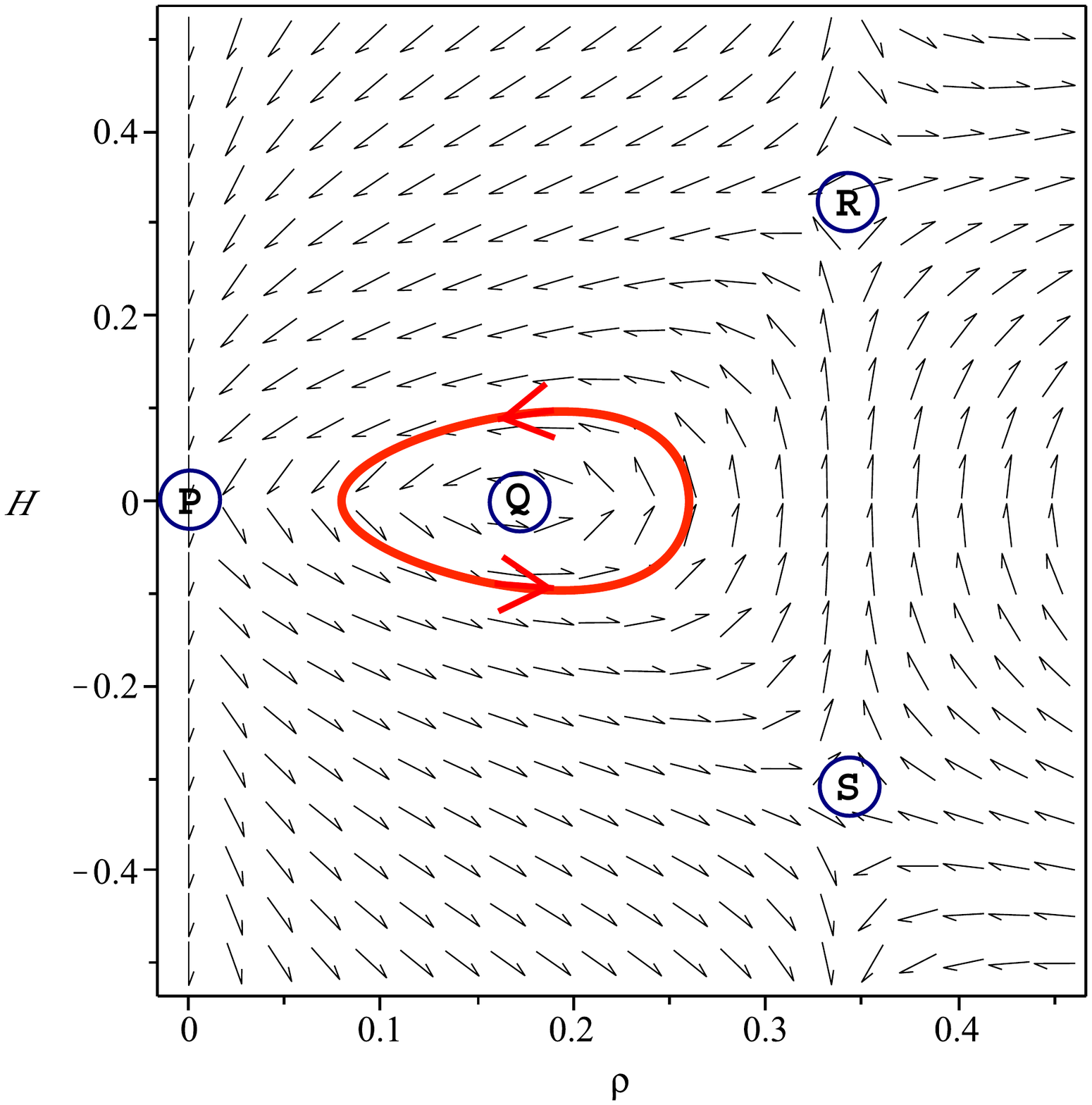}
\vskip-1.4cm
\noindent
\parbox{145mm}{Figure 3:  \footnotesize In linear approximation, the diagram on the phase plane is an ellipse around the centre $Q$. Points $R$
and $S$ are saddles, the origin $P$ attracts trajectories. This is  a model of a cyclic Universe.}
\vskip.4cm
\end{center}
\n
Very close to the equilibrium point $(\rho^\ast_2 = \rho_{\mbox{\tiny min}} \, , H^\ast_{1} = 0)$, i.e. when $r$ and $h$ are very small, the linear
approximation $\dot{r} = -3 [\alpha z(T)]^{-1} h, \quad \dot{h} = (T/2) r$ can be easily integrated to reveal that $r$ and $h$ sweep an ellipse in the
phase plane:
\b
\label{uno}
r & = & C \cos \omega t, \\
\label{due}
h & = & C Q \sin \omega t,
\e
where $C$ is a constant depending on the initial conditions,
\b
\omega = \sqrt{\frac{3 T}{2 \alpha  z(T) }}
\e
is the angular frequency of the oscillations and
\b
Q = \sqrt{ \frac{\alpha z(T) }{6}}
\e
is the ratio of the semi-axes of the ellipse. \\
It worth noticing that for the cooling Universe, with the drop of the temperature ($T \to 0$), the angular frequency $\omega$ decreases to zero and
in result the period of oscillations increases indefinitely and thus periodicity is lost, i.e. only the hot Universe is cyclic. At the same time, the ratio
$Q$ increases. \\
Unlike the case of the origin, where there was a doubly degenerated eigenvalue and the degeneracy was dealt with the introduction of the
infinitesimal perturbation parameter $\delta$, for point $(\rho^\ast_2 = \rho_{\mbox{\tiny min}} \, , H^\ast_{1} = 0)$ there is no degeneracy of the
spectrum of the stability matrix. The stability can be analyzed as follows. The system could be perturbed by variation of the parameters of the
model. The purely imaginary eigenvalue $\lambda$, determined in (\ref{i}), does not pick up, under such parameter perturbation, any real part for
the following reason. At point $(\rho^\ast_2 = \rho_{\mbox{\tiny min}} \, , H^\ast_{1} = 0)$, the trace of the stability matrix
$L( \rho_{\mbox{\tiny min}} \, , 0),$ as seen in (\ref{L}), is proportional to $H^\ast_1$ which does not depend on any parameters and is zero. Thus
any parameter variation will not alter the trace of the matrix $L( \rho_{\mbox{\tiny min}} \, , 0)$ and in result, the real part of  $\lambda$ will always
remain zero, as it is half of the trace of $L( \rho_{\mbox{\tiny min}} \, , 0)$. Therefore, the closed trajectories are preserved and stable, depending
on the initial conditions only. \\
It is also very interesting to study the passage of the trajectories on the phase plane through regions on the phase plane characterized by inflation.
By definition, inflation is equivalent to $\ddot{a}(t) > 0$ and $\dot{a}(t) > 0$. Since
\b
\frac{\ddot{a}}{a} = \dot{H} + H^2
\e
and since the scale factor $a(t)$ is always strictly positive, inflation occurs when $\dot{H} + H^2 > 0$. Upon substitution of the dynamical equation
(\ref{hash}), the condition for inflation becomes $H^2 < - p$ (the pressure $p$ must therefore be negative). Namely, the regions in the upper half [where
$H > 0$, i.e. $\dot{a}(t) > 0$] of the phase plane for which $H^2 < \alpha T z(T) \rho^2 - T \rho$ are {\it inflationary}. Or:
\b
H^2 < \alpha T z(T) \Bigl(\rho - \frac{\rho_{\mbox{\tiny min}}}{2}\Bigr)^2 - \frac{1}{4} \frac{T}{\alpha z(T)}.
\e
The boundary of this region is a hyperbola with asymptotes
\b
H = \pm \sqrt{\alpha T z(T)} \Bigl(\rho - \frac{\rho_{\mbox{\tiny min}}}{2}\Bigr)
\e
and cutting the $\rho$-axis exactly at $\rho = \rho_{\mbox{\tiny min}} = [\alpha z(T)]^{-1}$ (the left branch of the hyperbola is for negative values of
$\rho$ and thus it is of no interest) --- see Figure 4. With the drop of the temperature ($T \to 0$), the angle $CAK$ between the asymptotes
decreases to zero and the inflationary region disappears from the phase plane.
\begin{center}
\vskip-1.5cm
\includegraphics[width=10.5cm]{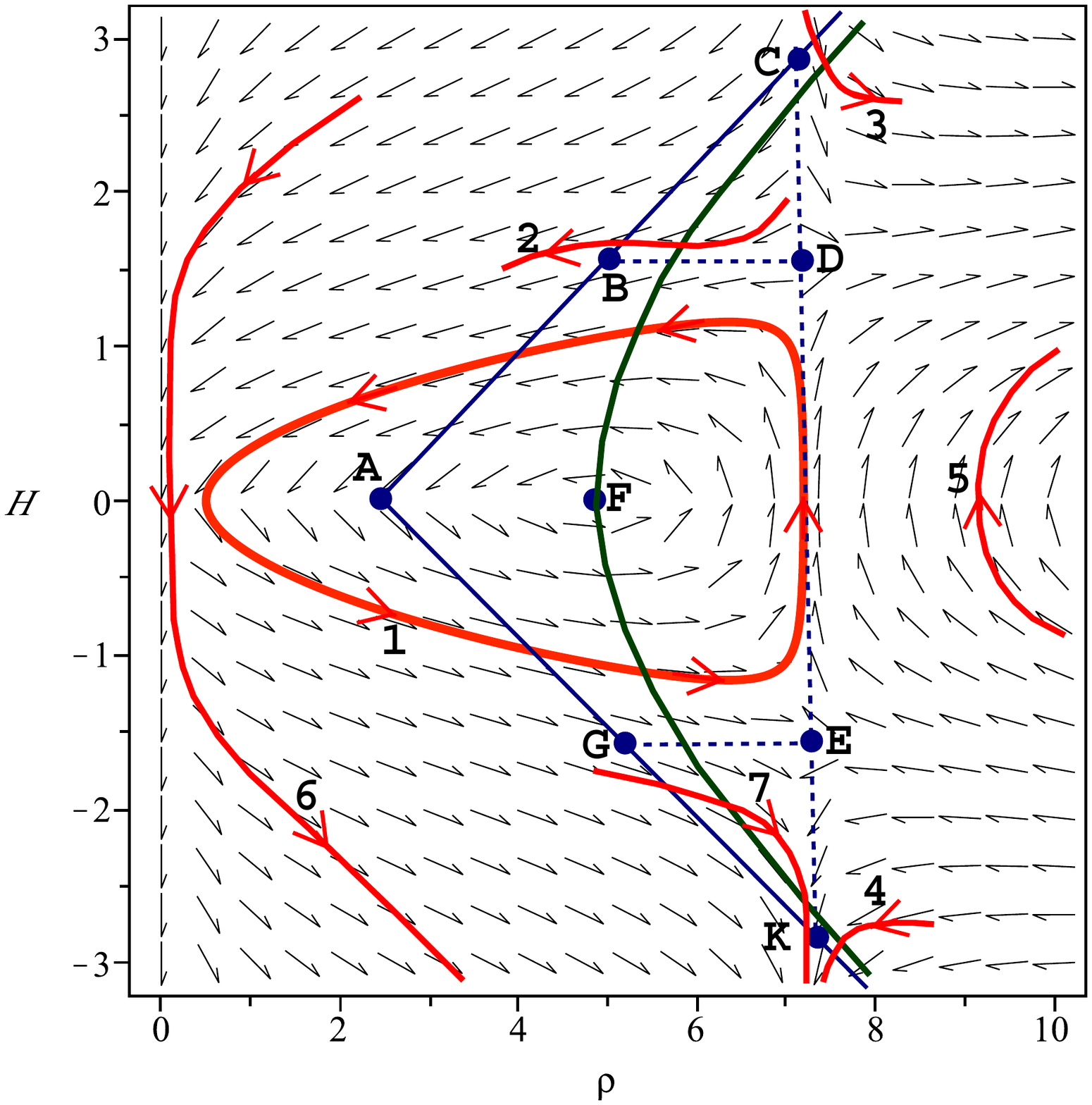}
\vskip-2.4cm
\noindent
\parbox{145mm}{\hskip0.83cm Figure 4:  \footnotesize The inflationary region is in the upper half-plane to the right of the hyperbola. }
\vskip.4cm
\end{center}
On Figure 4, the equilibrium points are as follows: the origin,  the centre $F$ with coordinates
$(\rho^\ast_2 = \rho_{\mbox{\tiny min}} \, , H^\ast_{1} = 0)$, the saddles $D$ and $E$ with coordinates $(\rho^\ast_3 \, , H^\ast_{2,3})$
respectively. \\
When $\rho = \rho^\ast_3 =  \rho_{\mbox{\tiny min}} (1 + 1/T)$, the asymptotes intersect the line $DE$ at points $C$ and $K$ with
$H$-coordinates given by $\pm \sqrt{\alpha T z(T)} \rho_{\mbox{\tiny min}} (1/2 + 1/T)$ respectively. The saddle points $D$ and $E$ are always
(i.e. for all $T$) between the asymptotes --- in the inflationary region (for $D$) and symmetrically below it (for $E$). \\
The cyclic Universe (trajectory 1 on Figure 4) experiences an inflationary period when the trajectory on the phase-plane is in the upper half-plane to the
right of the hyperbola. Of course, there is initial data that evolves into open trajectories on the phase plane that do not experience inflation at all
(trajectories 4, 6, 7), exit a region of initial inflation (trajectory 2), enter seemingly eternal inflation (trajectory 3) or stay in a seemingly eternal inflation
(trajectory 5). In the latter two cases however, with the drop of the temperature and the ``closure" of the inflationary region,  inflation will eventually end.
Obviously, as the hyperbola goes through the centre $F$, there is no initial data leading to inflation-only cyclic solutions. \\
To see how much inflation can be ``generated" in this model, it is useful to consider the $e$--fold number $N= \ln a$.
As $H = \dot{a}/a = (d/dt) \ln a$, the $e$--fold number $N$ is equal to $\int_{t_i}^{t_f} H(t) \, dt$. Using the linear approximation (\ref{due}), see Figure 3, results in
$N = \int_{t_i}^{t_f}  C Q \sin \omega t \, dt = (C Q / \omega) (\cos \omega t_i - \cos \omega t_f)$. During the inflationary phase, $\omega t_i =0$ and $\omega t_f$ is approximated by $\pi/2$. Thus the term in the last brackets is approximately 1. The upper limit of $CQ$ is the $H$--coordinate  of point $D$ on Figure 4, that is 
$H^\ast_{2}  =  \sqrt {\rho^\ast_3 / 3} =
\sqrt {\rho_{\mbox{\tiny min}} (1 + 1/T)/3} = \sqrt {(1 + 1/T)/[3 \alpha z(T)]}$. Thus, the upper limit of the $e$--fold number $N$ is $(\sqrt{2}/3) \sqrt{1/T + 1/T^2}$,
where $T$ is the temperature at the end of the inflationary phase.

\section{Van der Waals Gas Revisited}
\n
Many aspects of the van der Waals quintessence scenario have been investigated in a series of works of Capozziello {\it et al.} \cite{cap1},
\cite{cap2}, \cite{cap3}. The purpose of this section is to briefly recall the properties of the van der Waals quintessece model from the point of view
of dynamical systems and to compare to the case of virial real gas. \\
An interesting property of this model is the fact that the absolute temperature is again treated as a parameter, which can be negative. We will
retain this peculiarity only in view of the fact that a cyclic Universe possibility is realized when the absolute temperature varies over an entirely
negative interval. \\
The density of the Universe, $\rho$, will be treated as a strictly non-negative quantity. \\
In the dynamical equations, the energy density $\rho$ and the pressure $p$ are that of the van der Waals gas. The van der Waals equation of
state is \cite{mandl} (see also \cite{cap1}, \cite{cap2} and \cite{cap3}):
\b
\label{dw}
p = \frac{\g \rho}{1 - \beta \rho} - \alpha \rho^2 \, ,
\e
where $\alpha = 3 p_c/\rho_c^2$ and $\beta = 1/(3 \rho_c)$, with $\rho_c$ and $p_c$ being the density and pressure of the van der Waals gas at
the critical point. Here $\g$ is the temperature of the van der Waals gas. It will be treated as a parameter of the model \cite{cap1}, \cite{cap2},
\cite{cap3}. \\
In terms of a dimensionless energy density $\eta$, defined via $\eta = \rho/\rho_c$,  the van der Waals equation of state (\ref{dw}) becomes:
\b
\label{eeta}
p = \frac{\g \rho}{1 - \frac{\eta}{3}} - 3 p_c \eta^2 \, .
\e
At the critical point, $\rho = \rho_c$ (thus $\eta = 1$) and $p = p_c$. At this point, equation (\ref{eeta}) yileds $p_c = (3/8) \g \rho_c$. Therefore,
$\alpha = (27/8) \g \beta$ --- the three parameters $\alpha, \beta$, and $\g$ are not independent \cite{cap2} and only $\g$ and $\rho_c$  will be
treated as independent parameters.  Substituting $p_c = (3/8) \g \rho_c$ into (\ref{eeta}) yields \cite{cap2}:
\b
p = \frac{3 \g \rho}{3 - \eta} - \frac{9}{8} \g \eta \rho \, .
\e
For the parameter $\rho_c$ the following holds \cite{cap2}:
\b
\Omega_c = \frac{\rho_c}{\rho_{crit}} \, ,
\e
where the quintessence density parameter $\Omega_c$ is equal to $(1 - \Omega_{b, 0}) / \eta_0$ (here $\eta_0$ is the present day energy density of
the van der Waals gas and $\Omega_{b, 0}$  is the baryonic density parameter, given by $\rho_b(z=0)/\rho_{crit}$ with $z$ being the red-shift) and the
critical density of the Universe is $\rho_{crit} = 3 H_0^2/(8 \pi G)$ \cite{cap2}. \\
Substituting the van der Waals pressure into the dynamical equations (\ref{hash}) and (\ref{rho}) results in:
\b
\label{dyn2}
\dot{\eta} & = & - 3 H \eta \,  (1 + \frac{3 \g}{3 - \eta} - \frac{9}{8} \g \eta) \equiv g_1(\eta, H), \\
\label{dyn1}
\dot{H} & = & - \frac{3}{2} H^2 - \frac{8 \xi \g \eta }{3 - \eta} + 3 \xi \g \eta^2 \equiv g_2(\eta, H),
\e
where $\xi = (3/16)  \rho_c = $ const $>0$ is a parameter of the model (instead of the critical density $\rho_c$). \\
Next, the existence of equilibrium points will be addressed. Requesting $\dot{g}_1(\eta, H) = 0$, (\ref{dyn2}) yields the quadratic equation:
\b
\label{roots}
\frac{9}{8} \g \eta^2 - (\frac{27}{8} \g  + 1) \eta + 3(1 + \g) = 0 \, .
\e
For real roots to exist, the discriminant
\b
\label{discr}
D = - \frac{135}{64} \g^2 - \frac{27}{4} \g + 1
\e
must be positive. Thus, the range of values of $\g$, for which two real equilibrium points for $\eta$ exist is:
\b
\label{gamma}
-\frac{8}{5} - \frac{32}{45} \sqrt{6} < \g < -\frac{8}{5} + \frac{32}{45} \sqrt{6}
\e
or $-3.3419 < \g < 0.1419$. \\
Again, the physically meaningful range of $\g$ is: $0 < \g < -\frac{8}{5} + \frac{32}{45} \sqrt{6}$. \\
The corresponding two real values of $\eta$ are:
\b
\label{eta_roots}
\eta_{1, 2}^\ast(\g) = \frac{\frac{27}{8} \g  + 1 \pm \sqrt{- \frac{135}{64} \g^2 - \frac{27}{4} \g + 1 }}
{\frac{9}{4}\g} \, .
\e
Next, $\dot{g}_2(\eta, H) = 0$ results in four equilibrium values for $H$:
\b
\label{hash_roots}
H^\ast_{1, 2, 3, 4}(\g) = \pm \sqrt{\frac{2 \xi \eta_{1, 2}^\ast}{3}}\sqrt{\frac{\g (3 \eta_{1, 2}^{\ast^2} - 9 \eta_{1, 2}^\ast + 8)}{\eta_{1, 2}^\ast - 3}} \, .
\e
For these to be real, the following must hold:
\b
\g (\eta_{1, 2}^\ast - 3) (3 \eta_{1, 2}^{\ast^2} - 9 \eta_{1, 2}^\ast + 8) > 0 \, .
\e
As $3 \eta_{1, 2}^{\ast^2} - 9 \eta_{1, 2}^\ast + 8 = (\eta_{1, 2}^\ast - 3/2)^2 + 5/12 > 0,$  taking into consideration (\ref{gamma}), there are two
regimes:
\begin{itemize}
\item[\bf{(a)}] $-\frac{8}{5} - \frac{32}{45} \sqrt{6} < \g < 0$ and $\eta_{1,2}^\ast < 3$
\end{itemize}
\noindent
or
\begin{itemize}
\item[\bf{(b)}] $0 < \g < -\frac{8}{5} + \frac{32}{45} \sqrt{6}$ and $\eta_{1,2}^\ast > 3$
\end{itemize}
Indeed,  when $-\frac{8}{5} - \frac{32}{45} \sqrt{6} < \g < 0$, $\eta_{1, 2}^\ast$ is always less than 3 and when
$0 < \g < -\frac{8}{5} + \frac{32}{45} \sqrt{6}$, $\eta_{1, 2}^\ast$ is always greater than 3:
\addtocounter{figure}{+4}
\begin{figure}
\centering
\subfloat[$\eta_1^\ast(\g), -3.3419 < \g < 0.1419$]{\label{e1}\includegraphics[width=0.33\textwidth]{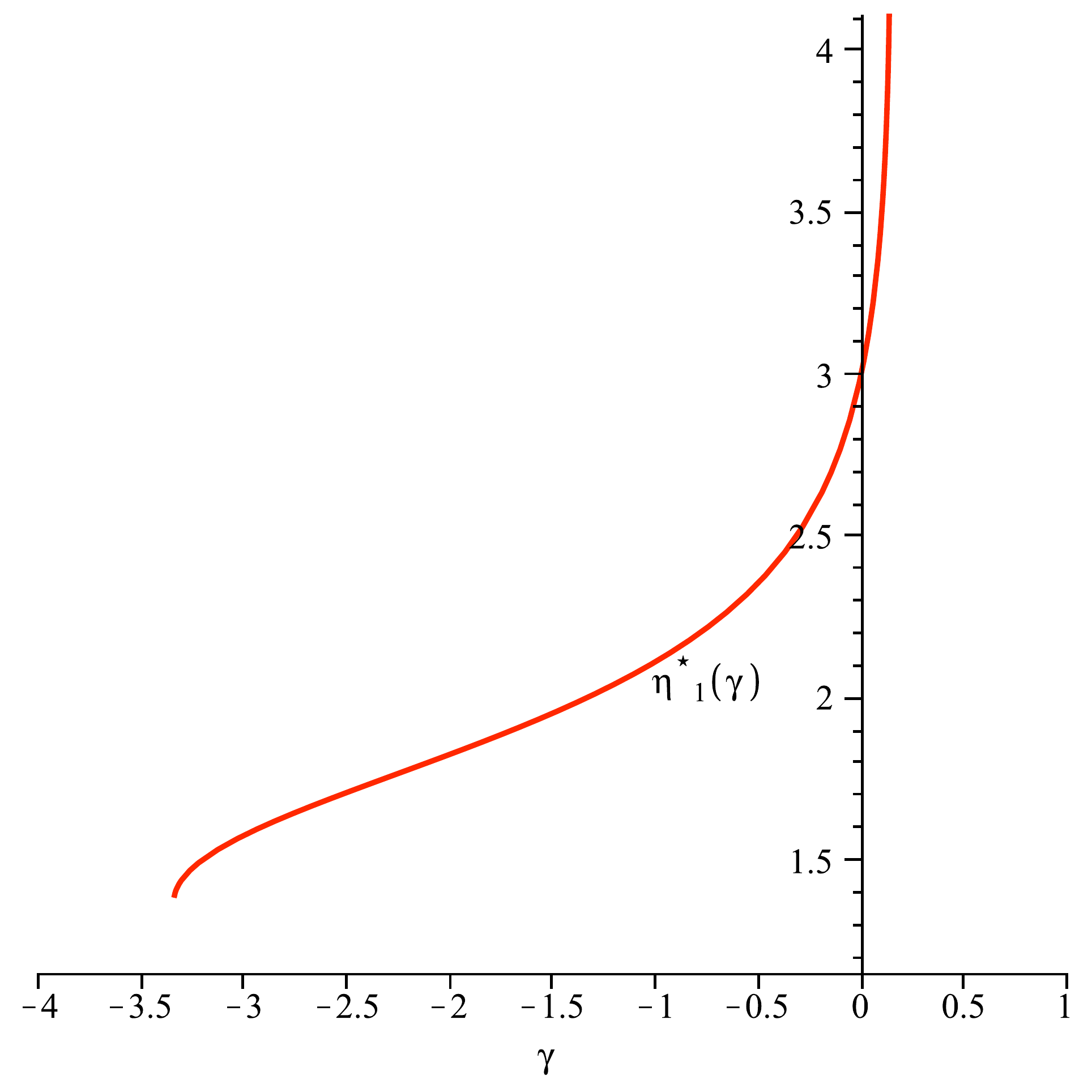}}
\subfloat[$\eta_2^\ast(\g), -3.3419 < \g < 0$]{\label{e2neg}\includegraphics[width=0.33\textwidth]{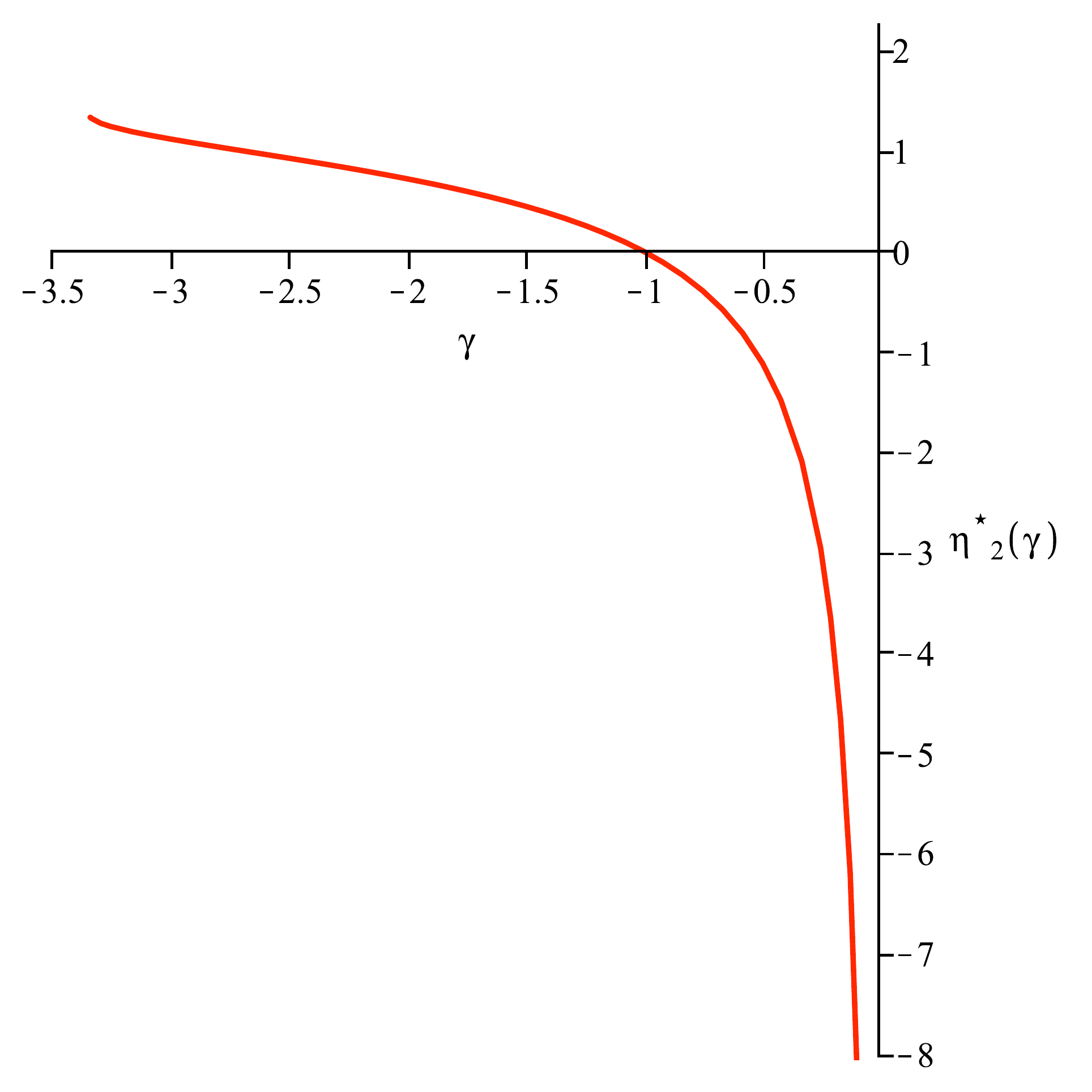}}
\subfloat[$\eta_2^\ast(\g), 0 < \g < 0.1419$]{\label{e2pos}\includegraphics[width=0.33\textwidth]{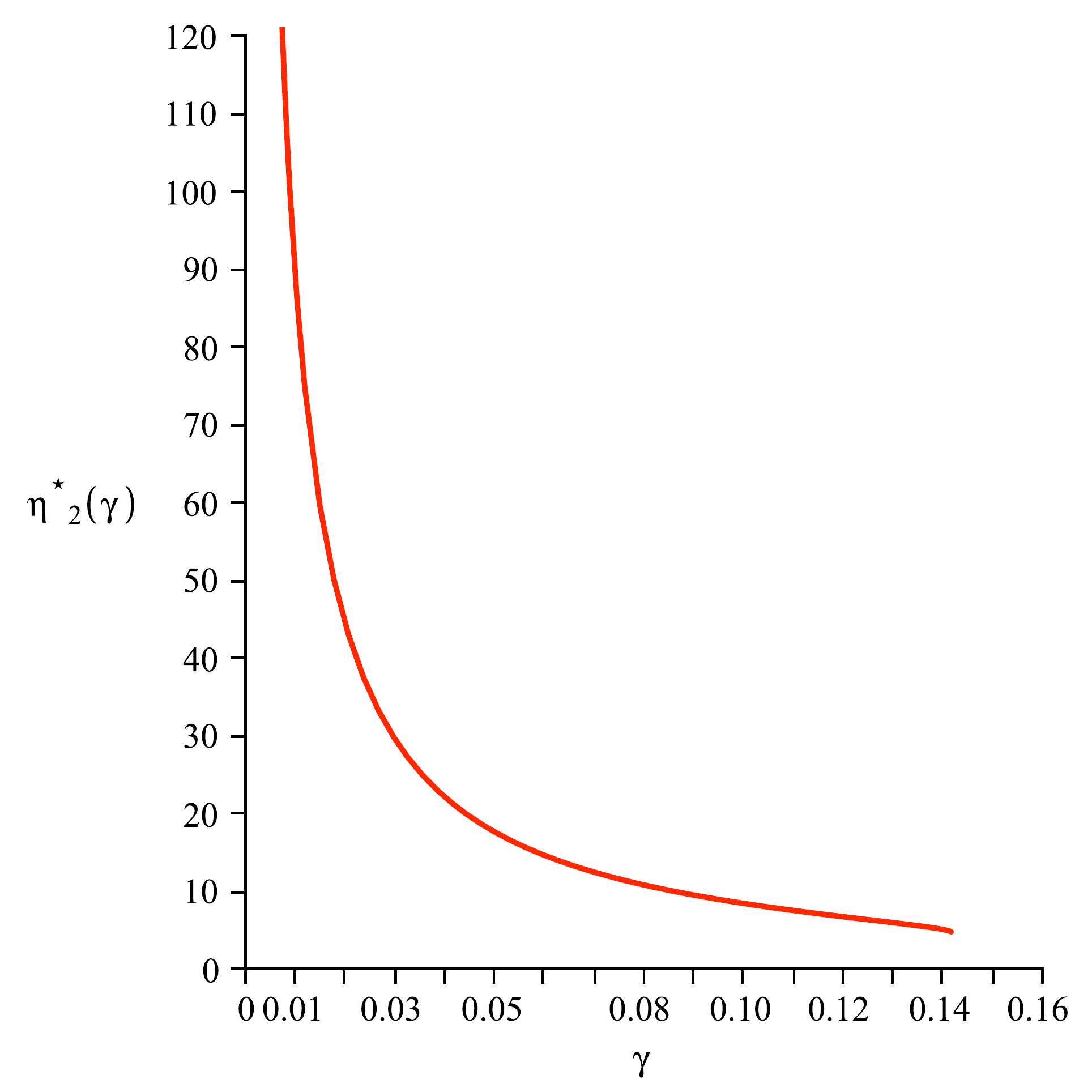}}
\caption{\footnotesize{The bigger root $\eta_1^\ast(\g)$ [on graph (a)] is a continuous function of $\g$, satisfying  $\eta_1^\ast(\g) < 3$ when
$-3.3419 < \g < 0$ and $\eta_1^\ast(\g) > 3$ when $0 < \g < 0.1419.$ The smaller root $\eta_2^\ast(\g)$ is not continuous [graphs (b) and (c)].
It also satisfies  $\eta_2^\ast(\g) < 3$ when $-3.3419 < \g < 0$ and $\eta_2^\ast(\g) > 3$ when $0 < \g < 0.1419$. However, in the range
$-1 < \g < 0$, $\eta_2^\ast(\g)$ becomes negative [see graph (b)] and, thus, unphysical. For this range of $\g$, there is only one equilibrium point:
$\eta_1^\ast(\g)$}.}
\label{fig1}
\end{figure}
However, for $-\frac{8}{5} - \frac{32}{45} \sqrt{6} < \g < -1$, there is only one positive root: $\eta_1^\ast(\g)$ (the bigger root). The other one
corresponds to a non-physical negative energy density. \\
Therefore, there are three intervals for $\g$, for which equilibrium solutions exist:
\begin{itemize}
\item[\bf{(1)}] $-\frac{8}{5} - \frac{32}{45} \sqrt{6} < \g < -1$ (i.e. $-3.3419 < \g < -1)$
\end{itemize}
There are two equilibrium values of $\eta$, namely $\eta_{1,2}^\ast(\gamma)$, and four corresponding equilibrium values of $H$, given by
(\ref{hash_roots}).
\begin{itemize}
\item[\bf{(2)}] $-1 < \g < 0$
\end{itemize}
There is only one positive equilibrium point $\eta_1^\ast(\g)$. The corresponding equilibrium values of $H$ are now two.
\begin{itemize}
\item[\bf{(3)}] $0 < \g < -\frac{8}{5} + \frac{32}{45} \sqrt{6}$ (i.e. $0 < \g < 0.1419$)
\end{itemize}
Again, there are two equilibrium values of $\eta$ and four equilibrium values of $H$. \\
At the equilibrium points $\eta_{1,2}^\ast(\g)$ and $H^\ast_{1,2,3,4}(\g)$, determined from (\ref{eta_roots}) and (\ref{hash_roots}), after setting
$g_1(\eta, H) = 0$ and $g_2(\eta, H) = 0$ in (\ref{dyn2}) and (\ref{dyn1}) respectively, the eigenvalues of the matrix $L(\eta, H)$ of the linearized
dynamical system are:
\b
\label{l1}
\lambda_1 & = & - 9 \g H^\ast_{1,2,3,4}(\g) \, \eta_{1,2}^\ast(\g) \, \biggl\{ \frac{1}{[3 - \eta_{1,2}^\ast(\g)]^2} - \frac{3}{8} \biggr\}, \\
\label{l2}
\lambda_2 & = & - 3H^\ast_{1,2,3,4}(\g).
\e
As the physically relevant $\eta_{1,2}^\ast(\g)$ are positive ($\eta$ is energy density), the signs of $\lambda_{1,2}$ are
determined by the sign of $\g$, the sign of $H^\ast_{1,2,3,4}(\g)$, and the sign of the expression in the figure brackets in (\ref{l1}). \\
The latter term, $[3 - \eta_{1,2}^\ast(\g)]^{-2} - 3/8,$ is positive when: $3 \eta_{1,2}^{\ast^2}(\g) - 18 \eta_{1,2}^\ast(\g) + 19 < 0$ or
$3 - (2/3) \sqrt{6} < \eta_{1,2}^\ast(\g) < 3 + (2/3) \sqrt{6}$. That is, $1.3670 < \eta_{1,2}^\ast(\g) < 4.6330$. \\
In the first interval of $\g$ for which equilibrium points exist,
\begin{itemize}
\item[\bf{(1)}] $-\frac{8}{5} - \frac{32}{45} \sqrt{6} < \g < -1$ (i.e. $-3.3419 < \g < -1)$,
\end{itemize}
there are four equilibrium points:
\begin{center}
\vskip1.4cm
\includegraphics[width=12cm]{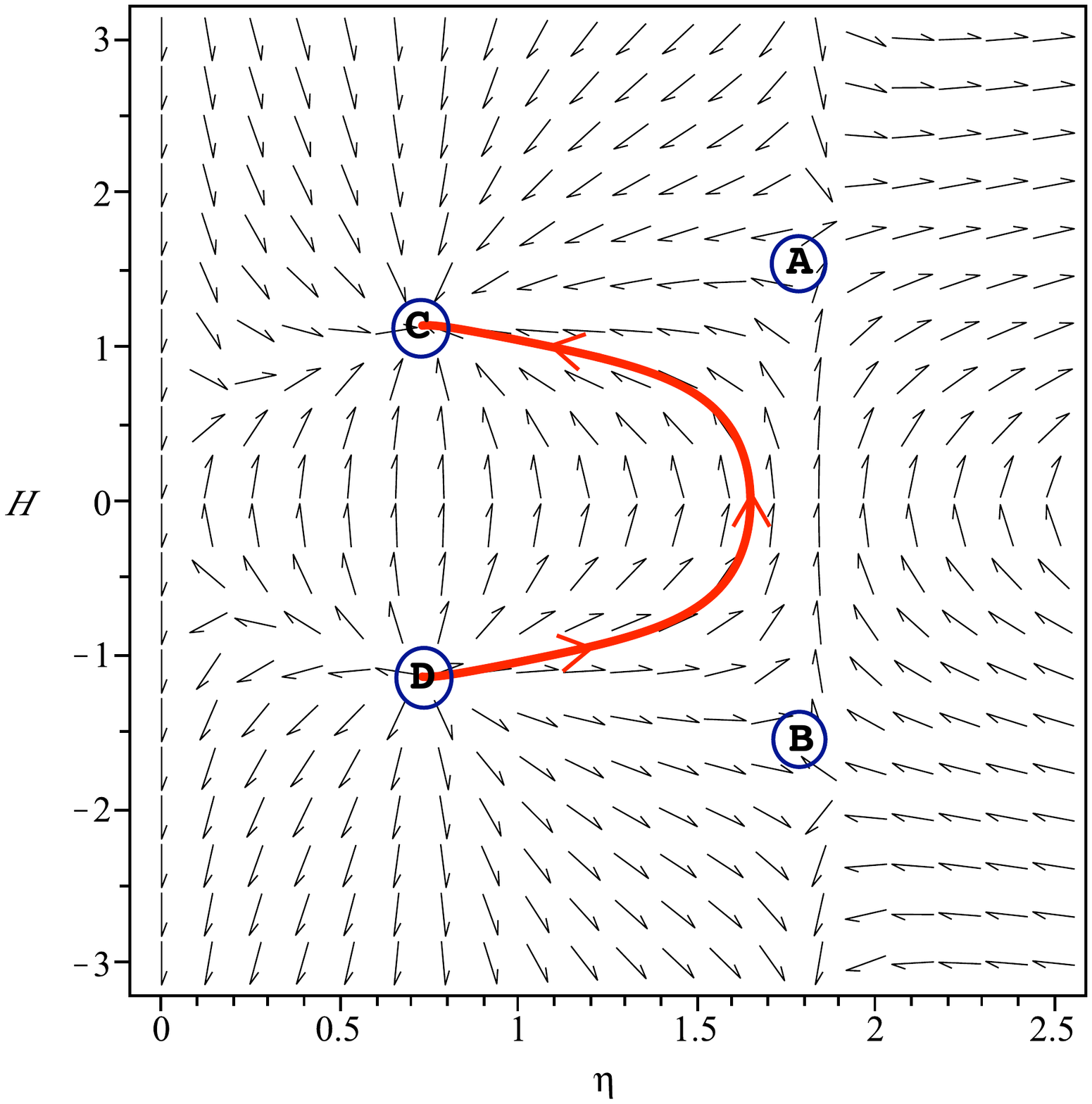}
\vskip.5cm
\noindent
\parbox{145mm}{Figure 6:  \footnotesize For $-3.3419 < \g < -1$ (and $\xi = 1$), points $A$ and $B$ are saddles, point $C$ --- stable
node, point $D$ --- unstable node.}
\end{center}
\vskip1cm
\begin{itemize}
\item[\bf{(i)}] Point $A$ with coordinates $H^\ast_1(\g) > 0$ and $1.3670 < \eta_{1}^\ast(\g) < 4.6330$. At this point $\lambda_1 > 0$ and
$\lambda_2 < 0$. This is a {\it saddle point}.
\item[\bf{(ii)}] Point $B$ with coordinates $H^\ast_2(\g) < 0$ and $1.3670 < \eta_{1}^\ast(\g) < 4.6330$. The eigenvalues flip signs:
$\lambda_1 < 0$ and $\lambda_2 > 0$. Another {\it saddle point}.
\item[\bf{(iii)}] Point $C$ with coordinates  $H^\ast_3(\g) > 0$ and $\eta_{2}^\ast(\g) < 1.3670$. At this point, both eigenvalues $\lambda_{1,2}$ are
negative and this is a {\it stable node}.
\item[\bf{(iv)}] Point $D$ with coordinates  $H^\ast_4(\g) < 0$ and $\eta_{2}^\ast(\g) < 1.3670$. At this point, both eigenvalues $\lambda_{1,2}$ are
positive, corresponding to an {\it unstable node}.
\end{itemize}
In the next interval of $\g$ with equilibrium points:
\begin{itemize}
\item[\bf{(2)}] $-1 < \g < 0$,
\end{itemize}
there are only two equilibrium points (recall that $\eta_2^\ast(\g)$ is negative in that range):
\begin{center}
\vskip1.4cm
\includegraphics[width=12cm]{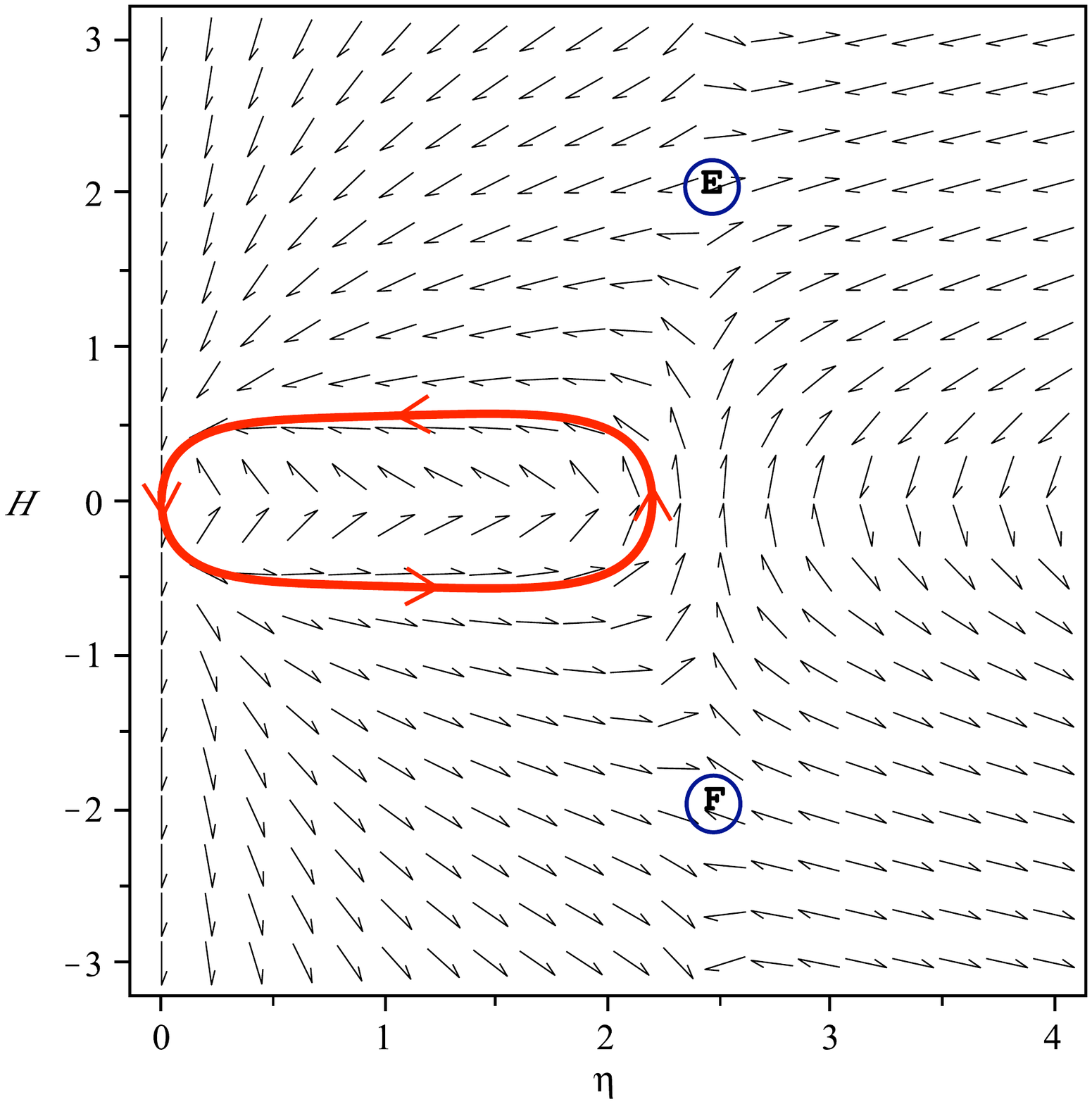}
\noindent
\vskip.5cm
\noindent
\parbox{145mm}{Figure 7:  \footnotesize For $-1 < \g < 0$ (and $\xi = 1$), there are only two equilibrium points --- $E$ and $F$ --- both saddles.}
\end{center}
\vskip1cm
\begin{itemize}
\item[\bf{(i)}] Point $E$ with coordinates $H^\ast_1(\g) > 0$ and $1.3670 < \eta_{1}^\ast(\g) < 4.6330$. At this point $\lambda_1 > 0$ and
$\lambda_2 < 0$ --- {\it saddle point}.
\item[\bf{(ii)}] Point $F$ with coordinates $H^\ast_2(\g) < 0$ and $1.3670 < \eta_{1}^\ast(\g) < 4.6330$. The eigenvalues flip signs:
$\lambda_1 < 0$ and $\lambda_2 > 0$ --- {\it saddle point}.
\end{itemize}
This is again a cyclic Universe (but corresponding to a toy model with negative absolute temperature). \\
Finally, in the last interval of $\g$ with equilibrium points:
\begin{itemize}
\item[\bf{(3)}] $0 < \g <  -\frac{8}{5} + \frac{32}{45} \sqrt{6}$ (i.e. $0 < \g < 0.1419)$,
\end{itemize}
there are again four equilibrium points:
\begin{center}
\includegraphics[width=12cm]{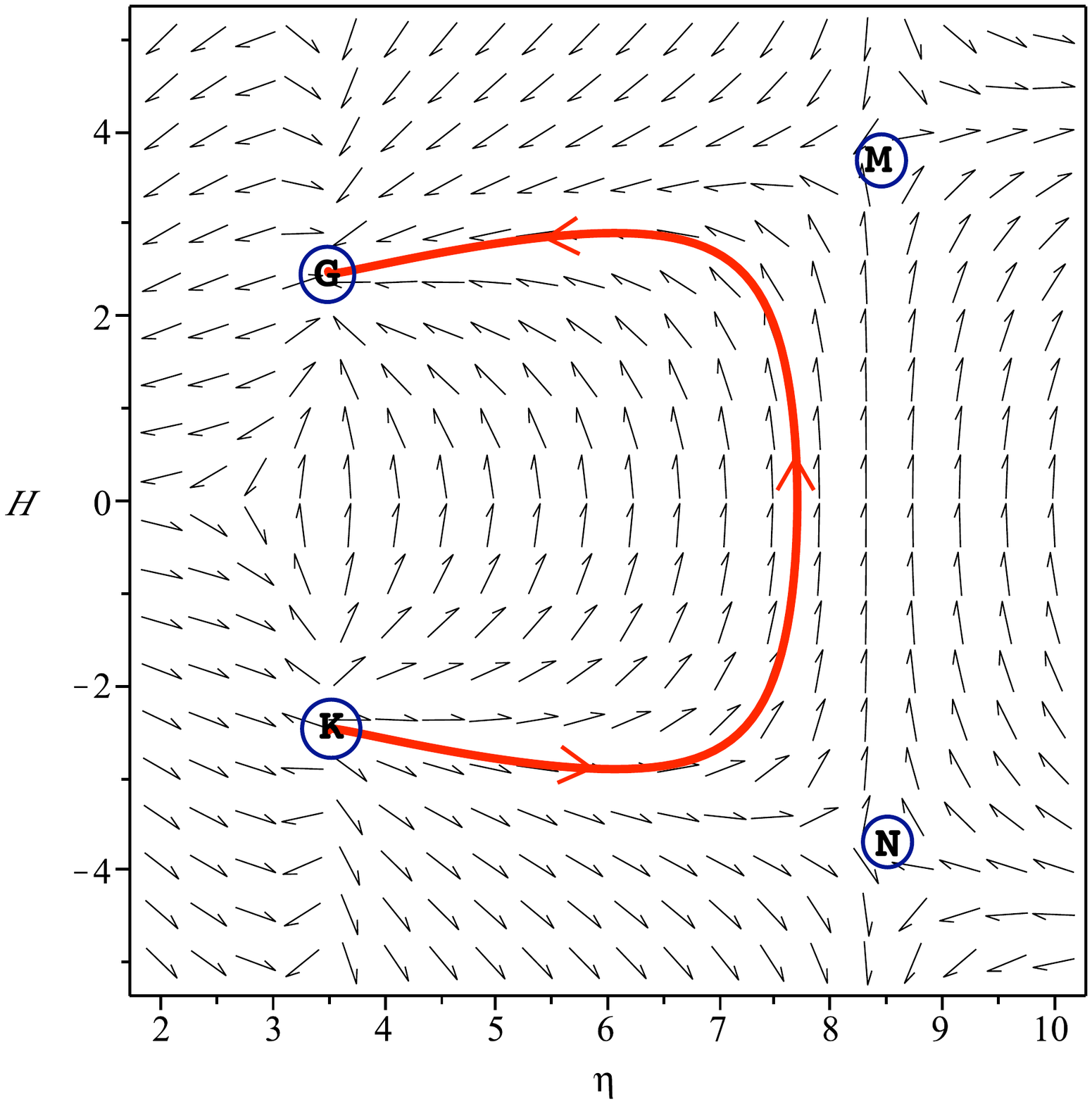}
\vskip.6cm
\noindent
\parbox{145mm}{Figure 8:  \footnotesize For $0 < \g < 0.1419$ (and $\xi = 1$), points $M$ and $N$ are saddles, point $G$ --- stable node, point
$K$ --- unstable node.}
\end{center}
\begin{itemize}
\item[\bf{(i)}] Point $G$ with coordinates $H^\ast_1(\g) > 0$ and $1.3670 < \eta_{2}^\ast(\g) < 4.6330$. Both eigenvalues $\lambda_{1,2}$ are
negative and this is a {\it stable node}.
\item[\bf{(ii)}] Point $K$ with coordinates $H^\ast_2(\g) < 0$ and $1.3670 < \eta_{2}^\ast(\g) < 4.6330$. Both eigenvalues $\lambda_{1,2}$ are
positive and this is an {\it unstable node}.
\item[\bf{(iii)}] Point $M$ with coordinates  $H^\ast_3(\g) > 0$ and $4.6330 < \eta_{1}^\ast(\g)$. At this point $\lambda_1 > 0$ and
$\lambda_2 < 0$. This is a {\it saddle point}.
\item[\bf{(iv)}] Point $N$ with coordinates  $H^\ast_4(\g) < 0$ and $4.6330 < \eta_{1}^\ast(\g)$. The eigenvalues flip signs: $\lambda_1 < 0$ and
$\lambda_2 > 0$. Again, a {\it saddle point}.
\end{itemize}
The point $(\eta^\ast = 0, H^\ast = 0)$ in the phase portrait is also an equilibrium point. Apparently, at this point, $\lambda_1 = 0 = \lambda_2$ is a
double eigenvalue of the stability matrix $L(0, 0)$ --- as in the case of the virial real gas. Again, the introduction of an infinitesimal
parameter $\delta > 0$ in the same way in $H$ perturbs the dynamical system and bifurcates the origin into a ``dipole" of two equilibrium points.
We consider:
\b
\label{dy2}
\dot{\eta} & = & - 3 H \eta \,  (1 + \frac{3 \g}{3 - \eta} - \frac{9}{8} \g \eta) \, , \\
\label{dy1}
\dot{H} & = & - \frac{3}{2} H^2 + \frac{3}{2} \delta^2 - \frac{8 \xi \g \eta }{3 - \eta} + 3 \xi \g \eta^2 \, .
\e
Clearly, $(\eta_3^\ast = 0, H^{' \ast}_{5} = + \delta)$ and $(\eta_3^\ast = 0, H^{'' \ast}_{5} = - \delta)$ are the two equilibrium points of the ``dipole".
At these points, the eigenvalues of the stability matrix $L(0, \pm \delta)$ are:
\b
\label{le1}
\lambda_1 & = & - 3(\pm \delta)(1 + \g), \\
\label{le2}
\lambda_2 & = &  - 3(\pm \delta).
\e
There are only two possible cases --- the situation where $\g$ is positive (which leads to equilibrium points with $\eta^\ast $ greater than 3)
cannot be studied at the origin.\\
Firstly, if $-1 < \g < 0$, then  $\lambda_1$ and $\lambda_2$ have the same signs.  Thus, point $(\eta_3^\ast = 0, H_{5}^{' \ast} = + \delta)$  is a
{\it stable node}, while point $(\eta_3^\ast = 0, H_{5}^{'' \ast} = - \delta)$ is an {\it unstable node}. In the limit $\delta \to 0$, the two points coalesce
and this reveals that the origin attracts trajectories in the phase plane. The graph on Figure 7 is realized. \\
The other possible case is $-3.3419 < \g < -1$. Then  $\lambda_1$ and $\lambda_2$ have the different signs and the two equilibrium points of the
``dipole"  are {\it saddle points}. This shows that, upon taking the limit $\delta \to 0$, the origin will not be able attract any
trajectories in the phase plane.

\section{Conclusions}
It has been shown that there is a possible cyclic Universe scenario with inflation realizable within a classical physics model which offers a graceful
exit from inflation. The cyclic solution is possible for a wide range of initial data and a wide range of temperatures for real gas with equation of
state derived from the virial expansion and for negative temperatures for a van der Waals real gas. Periodicity is lost too as the temperature of the
Universe tends to zero.

\section*{Acknowledgments}
R.I.I. gratefully acknowledges the support of Science Foundation Ireland (SFI), grant No. 09/RFP/MTH2144.


\begin{thebibliography}{99}

\bibitem{1} S. Perlmutter {\it et al.}, {\it Measurements of the Cosmological Parameters Omega and Lambda from the First Seven Supernov\ae at
$z \ge 0.35$},  Astrophys. J. {\bf 483(1)}, 565--581 (1997); \\
S. Perlmutter {\it et al.}. {\it Discovery of a Supernova Explosion at Half the Age of the Universe}, Nature {\bf 391}, 51--54 (1998); \\
S. Perlmutter {\it et al.}, {\it Measurements of Omega and Lambda from 42 High-Redshift Supernov\ae}, Astrophys. J. {\bf 517(2)}, 565--586
(1999); \\
B.P. Schmidt {\it et al.}, {\it The High-Z Supernova Search: Measuring Cosmic Deceleration and Global Curvature of the Universe Using Type Ia
Supernov\ae}, Astrophys. J. {\bf 507(1)}, 46--63 (1998); \\
A.G. Riess {\it et al.}, {\it Observational Evidence from Supernovae for an Accelerating Universe and a Cosmological Constant}, Astronom. J.
{\bf 116(3)}, 1009--1038 (1998).

\bibitem{2} A.R. Liddle and D.H. Lyth, {\it Cosmological Inflation and Large-Scale Structure}, Cambridge University Press (2000).

\bibitem{3} R.R. Caldwell, R. Dave, and P.J. Steinhardt, {\it Cosmological Imprint of an Energy Component with General Equation of State;}
Phys. Rev. Lett. {\bf 80}, 1582--1585 (1998).

\bibitem{4} J.E. Lidsey and D.J. Mulryne, {\it Graceful Entrance to Braneworld Inflation},  Phys. Rev. {\bf D 73}, 083508 (2006).

\bibitem{5} D.J. Mulryne, N.J.  Nunes, R. Tavakol,  and J.E. Lidsey, {\it Inflationary Cosmology and Oscillating Universes in Loop Quantum
Cosmology}, Int. J. Mod. Phys. {\bf A 20(11)}, 2347--2357 (2005).

\bibitem{6} J.E. Lidsey, {\it Triality between Inflation, Cyclic, and Phantom Cosmologies}, Phys. Rev. {\bf D 70}, 041302 (2004).

\bibitem{7} S. Carloni, E. Elizalde, and P.J. Silva, {\it An Analysis of the Phase Space of Ho\^rava-Lifshits Cosmologies}, Class. Quant. Grav.
{\bf 27(4)}, 045004 (2010).

\bibitem{8} E.J. Copeland, S. Mizuno, and M. Shaeri, {\it Dynamics of a Scalar Field in Robertson-Walker Spacetimes}, Phys. Rev. {\bf D 79},
103515 (2009).

\bibitem{9} E.J. Copeland, A.R. Liddle, and D. Wands, {\it Exponential Potentials and Cosmological Scaling Solutions},  Phys. Rev. {\bf D 57},
4686--4690, (1998).

\bibitem{10} L. Parisi and R. Canonico, {\it  Modified Cosmological Equations and the Einstein Static Universe}, J. Geom. Symmetry Phys. {\bf 22},
51--65 (2011).

\bibitem{cap1} S. Capozziello, S. De Martino, and M. Falanga, {\it Van der Waals Quintessence}, Phys. Lett. {\bf A 299}, 494--498 (2002).

\bibitem{cap2}  S. Capozziello, V.F. Cardone, S. Carloni, S. De Martino, M. Falanga, A. Troisi, and M. Bruni, {\it Constraining van der Waals
Quintessence by Observations}, JCAP {\bf 04}, 005 (2005).

\bibitem{cap3} S. Capozziello, S. Carloni, A. Troisi, {\it Quintessence without Scalar Fields}, Recent Res. Dev. Astron. Astrophys. {\bf 1}, 625
(2003).

\bibitem{frwl} A. Friedmann, {\it On the curvature of space}, Gen. Rel. Grav. {\bf 31}, 1991 (1999) [Zeitschrift f\"ur Physik A {\bf 10}, 377--386 (1922)];
A. Friedmann, {\it On the possibility of a world with constant negative curvature of space}, Gen. Rel. Grav. {\bf 31}, 2001 (1999) [Zeitschrift f\"ur
Physik A{\bf 21}, 326--332 (1924)]; \\
G. Lema\^itre, {\it Un univers homog\`ene de masse constante et de rayon croissant, rendant compte de la vitesse radiale des n\`ebuleuses
extragalactiques,} Annales de la Soci\'et\`e Scientfique de Bruxelles A{\bf 47}, 49--56 (1927) [{\it A Homogeneous Universe of Constant Mass and
Increasing Radius Accounting for the Radial Velocity of Extra-Galactic Nebul\ae}, Mon. Not. R. Astr. Soc. {\bf 91}, 483--490 (1931)]; \\
H.P. Robertson, {\it Kinematics and World Structure, I}, Astrophys. J. {\bf 82}, 284--301 (1935); \\
A.G. Walker, {\it On the Formal Comparison of Milne�s Kinematical System with the Systems of General Relativity}, Mon. Not. Roy. Astr. Soc.
{\bf 95}, 263--269 (1935).

\bibitem{friedmann} P.J.A. Peebles, {\it Principles of Physical Cosmology}, Princeton University Press (1993).

\bibitem{mandl} L.D. Landau and E.M. Lifshitz, {\it Course of Theoretical Physics, Volume 5 --- Statistical Physics, Part 1}, Butterworth--Heinemann
(1980); \\
F. Mandl, {\it Statistical Physics}, Wiley (1982).

\bibitem{kih} T. Kihara, {\it Virial Coefficients and Models of Molecules in Gases}, Rev. Mod. Phys. {\bf 25(4)}, 831--843 (1953).

\end{thebibliography}
\end{document}